# Applications of Physics to Archery

H. O. Meyer

*Physics Department, Indiana University*

meyer1@indiana.edu

(6 November 2015)

Abstract:
Archery lends itself to scientific analysis. In this paper we discuss physics laws that relate to the mechanics of bow and arrow, to the shooting process and to the flight of the arrow. In parallel, we describe experiments that address these laws. The detailed results of these measurements, performed with a specific bow and arrow, provide insight into many aspects of archery and illustrate the importance of quantitative information in the scientific process. Most of the proposed experiments use only modest tools and can be carried out by archers with their own equipment.

## 1 INTRODUCTION

For more than 10,000 years, human civilizations relied on bow and arrow to provide food and to fight wars. As gunpowder gradually displaced human-powered weapons, archery declined until the 18[th] century when it experienced a revival as a recreational activity and as a modern sport.

The behavior of bows and arrows, the shooting process, and the flight of the arrow towards the target are described and explained to a large extent by physics, mainly mechanics, elasticity and aerodynamics. Recognizing this, bowmen with scientific knowledge began to perform quantitative experiments with their bows around 1920 [1,2]. An anthology of early scientific archery papers was published as a book in 1947 [3]. Insight gained from these studies questioned the traditional longbow design and transformed bow making from a craft to a science.

The continuing advancement of archery equipment, based on scientific principles, has resulted in the modern Olympic recurve bow and in the compound bow, which uses a system of cables and pulleys to modify the draw force. Crucial improvements are also due to the emergence of new plastics and compound materials, replacing traditional ingredients, such as wood, linen and animal hide.

This paper, written for scientists with an interest in archery, contains a discussion of physics laws that apply to various aspects of archery and a description of experiments to test these laws. Most of these measurements require only modest tools and can be performed by readers using their own equipment. In the context of this paper, data from the proposed experiments are collected for a specific bow (a compound bow) and a specific arrow, demonstrating how the understanding of many aspects of archery requires quantitative information.

## 2 THE ARROW

### 2.1 Shape, straightness and mass

A modern archery arrow is shown in Fig. 1. It consists of four parts: (1) a *shaft* made from tubular carbon-fiber compound, aluminum, or a combination thereof, (2) the *tip*, or 'pile' made from steel or brass, sometimes screwed into an aluminum insert, (3) *fletching*, consisting of three or four *fins* of a variety of materials, shapes and sizes, and (4) a *nock* at the rear end of the arrow, which clips onto the bowstring and is typically made of plastic and often mounted in an aluminum insert as well.

By convention, the length $L$ of an arrow is defined as the distance from the nocking point (where the bowstring touches the arrow) to the front end of the *shaft*, excluding the tip. We assume that the arrow is symmetric around the $z$-axis of a Cartesian frame. The nocking point fixes $z = 0$. The $y$-axis shall be up and the $x$-axis sideways. The shaft of the arrow consists of a hollow cylinder with an outer radius $R$ and wall thickness $\Delta R$.

The straightness of today's carbon arrows is excellent. For instance, the axis of a moderately-priced carbon arrow is guaranteed to deviate by less than 100 μm (about the thickness of human hair) from a straight line. One can test this by rotating an arrow resting on V-notches at both ends. The wobble amplitude of the shaft in the middle, observed with a microscope, equals the deviation from straight. We have measured 10 of our sample arrows and found deviations ranging from 10 to 100 μm, with an average of 55 μm, and an accuracy of about 5 μm. This is nice to know, but it is not clear if and how such a small deviation from straight will affect the trajectory of an arrow, especially when this arrow is rotating and oscillating.

A measurement of the mass $M$ of the arrow and that of its components requires a scale with a precision of ± 0.01 g. The distribution of the mass along the sample arrow, evaluated by weighing all parts separately, is shown in Fig. 1. From these data, the location $z_{cm}$ of the center of mass can be calculated. It agrees with the center of mass found by balancing the arrow on an edge.

A quantity called '*FOC*' is widely used to quantify the amount by which the center of mass is 'Forward-Of-Center', defined as

$$FOC \equiv \frac{z_{cm}}{L} - \tfrac{1}{2} \quad . \tag{1}$$

It is generally accepted by the archer's community that the *FOC* value should range from 0.07 to 0.17, but quantitative evidence backing up this belief seems to be lacking (see the tests with varying *FOC*, reported by Hickman in Ref. 3, p.77)

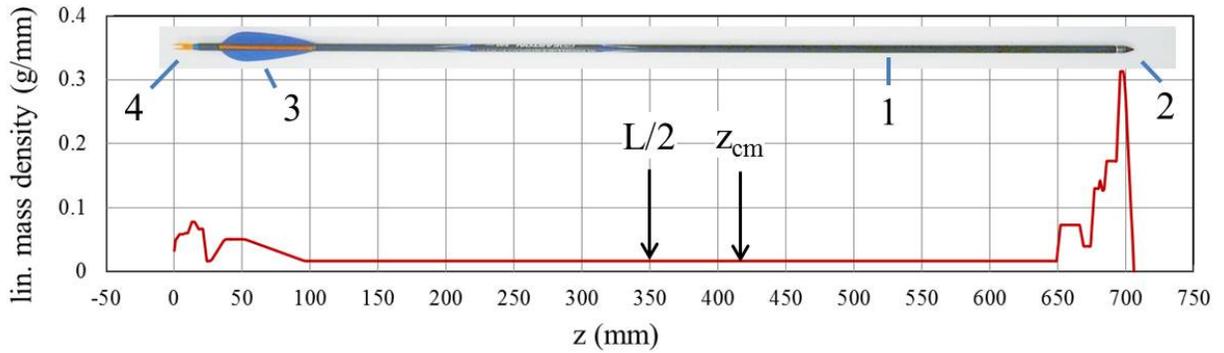

**Fig. 1** Linear mass distribution of the sample arrow [4]. The four parts of an arrow are indicated by numbers (see text)

Table 1. Static parameters of the sample arrow [4]

| | | | |
|---|---|---|---|
| $L$ | 0.696±0.001 | m | arrow length |
| $R$ | 3.63±0.05 | mm | shaft radius |
| $\Delta R$ | 0.50±0.07 | mm | wall thickness |
| $M$ | 20.62±0.01 | g | total arrow mass |
| $m_{shaft}$ | 11.70±0.01 | g | (57% of total) |
| $m_{tip}$ | 6.20±0.01 | g | (30%) |
| $m_{fins}$ | 1.41±0.01 | g | (7%) |
| $m_{nock}$ | 1.24±0.01 | g | (6%) |
| $z_{cm}$ | 0.414±0.001 | m | center of mass coordinate |
| $FOC$ | 0.093±0.002 | -- | 'Forward-Of-Center' |
| $S$ | 5.17±0.03 | Nm$^2$ | stiffness |
| $Spine$ | 492±3 | -- | ATA 'spine' |

## 2.2 Stiffness

The stiffness $S$ is a property of the shaft material and quantifies the ability of an arrow to bend. It is of crucial importance in archery for two reasons. First, the stiffness must have a *minimum* value to prevent destruction of the arrow during acceleration (sects. 4.1, 4.2), and second, it has to have a *specific* value for the proper interaction between the bow and the arrow during the shooting process (sect. 4.3).

The stiffness is easily measured as follows. In the setup shown in Fig. 2, a section of the arrow is supported by two knife edges, which allow the arrow to tilt around the *x*-direction. Small notches in the edges keep the arrow from rolling off sideways. The distance $\Delta z$ between the two supports is arbitrary, but must be known (here, $\Delta z$ = 0.550 m).

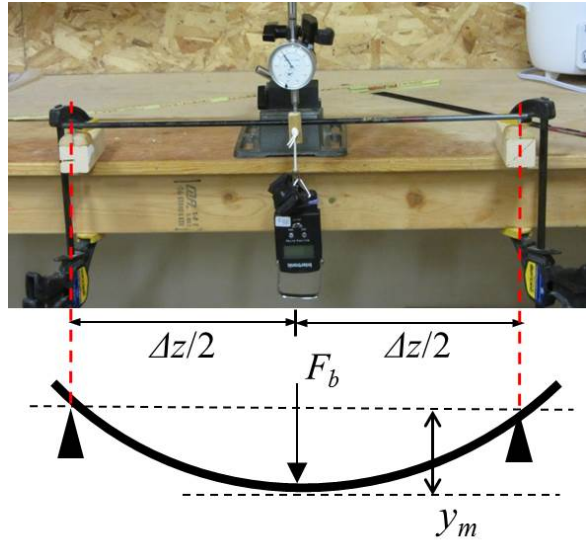

**Fig. 2** Setup to measure the stiffness of the sample arrow

A varying downward force $F_b$ is applied halfway between the supports by hanging weights or by pulling with a luggage scale. The downward deflection $y_m$ at the mid-point is measured with a dial indicator. The measured $y_m$ as a function of $F_b$ are shown in Fig. 3. As expected for elastic deformation, the deflection is proportional to the applied force,. The physics that governs the slope $y_m/F_b$ is understood within the Euler-Bernoulli beam theory [5] which is valid for small deflections of thin beams under lateral loads. The expected result for the present situation is

$$y_m \equiv F_b \frac{\Delta z^3}{48 S} \quad . \tag{2}$$

This equation defines the *stiffness* $S$ [Nm$^2$]. Since all other quantities in Eq. (2) have been measured, the stiffness can be deduced (listed in table 1).

The Archery Trade Association (ATA) proposes to measure stiffness by a different quantity, called *Spine* which is in common use. Their definition reads: "*Spine* is the number of thousands of an inch by which the center of an arrow shaft of 28 inch length is displaced when a sideways force of 1.94 pounds is acting at that point". To convert stiffness $S$ into *Spine*, the above definition, expressed in SI units, is inserted into Eq. 2. It follows that "*Spine* equals 2544 divided by the stiffness $S$ in Nm$^2$". The result for our sample arrow is 492, close to the value of *Spine* 500, quoted by the manufacturer.

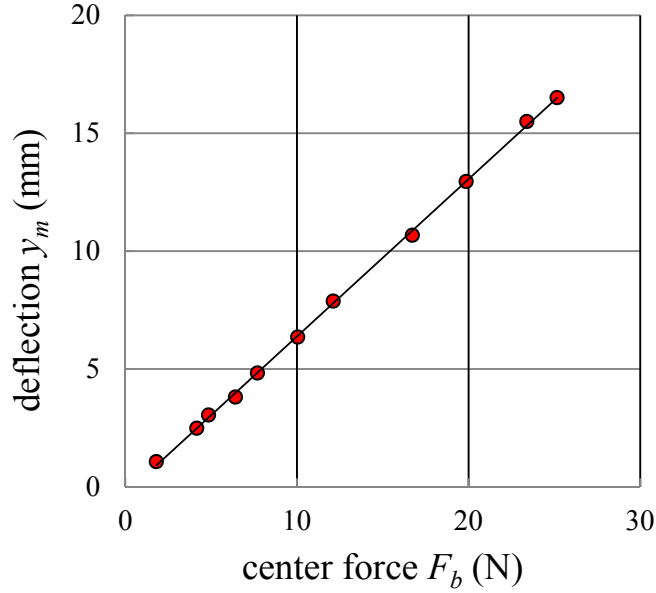

**Fig. 3** Deflection $y_m$ versus applied force $F_b$, measured with the setup shown in Fig.2

It turns out that stiffness is the product of two factors, one related to the *geometry* and the other to the *material properties* of the shaft.

$$S = J_{xx} \cdot Y_z \quad . \tag{3}$$

The geometry factor, $J_{xx}$, is the second moment of the beam cross-section relative to the *x*-axis. For a hollow cylinder with an outer radius of $R$ and inner radius of $R-\Delta R$, $J_{xx}$ is given by

$$J_{xx} = \int y^2 dx\, dy = \frac{\pi}{4}\left(R^4 - (R-\Delta R)^4\right) \quad . \tag{4}$$

With $R$ and $\Delta R$ from table 1, one obtains $J_{xx} = 6.10 \cdot 10^{-11}$ m$^4$. When we assume that $\Delta R$ is much smaller than $R$, the moment $J_{xx}$ (and thus the stiffness) is proportional to $\Delta R$ and to $R^3$. The stiffness of an arrow is therefore affected more strongly by the diameter than by the wall thickness.

The other factor in Eq. (3) is the elastic modulus, which quantifies strain in response to a given stress and is a material property. The mechanical properties of carbon fiber composite materials are not well defined and even depend on direction (here we need the modulus $Y_z$ that applies in the *z*-direction). However, we may treat $Y_z$ in Eq.(3) as the unknown and deduce from our data a value of $Y_z = 8.48 \cdot 10^{10}$ N/m$^2$, which is in line with literature values for carbon fiber materials.

## 2.3 Transverse oscillation

### 2.3.1 Measurement of the fundamental transverse oscillation

An arrow in flight is oscillating sideways, to some degree. The dominant 'fundamental' mode is shown in Fig. 4. The two stationary points $\alpha$ and $\beta$ are called *nodes*. Their positions ($z_\alpha = 0.125$ m, $z_\beta = 0.630$ m) can be calculated from a model as discussed in sect. 2.3.2.

An experiment to study the fundamental mode is described in the following. As shown in Fig. 5, the arrow is supported at the nodes by mounts that constrain the arrow vertically but allow tilting around a horizontal axis (Figs. 6a and 6b). Thus, the mechanical support does not interfere with the fundamental oscillation.

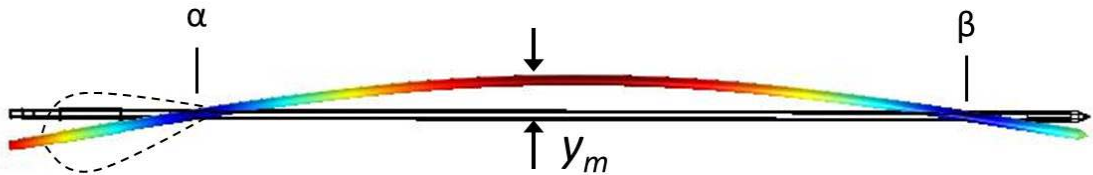

**Fig. 4** Fundamental mode of transverse oscillation. The coloration is related to displacement. The two nodes are labelled $\alpha$ and $\beta$

The oscillation is driven by a magnetic force acting on the steel tip. The driving magnet ('2' in Fig. 5) consists of a 200-turn coil on a 20 by 20 mm$^2$ iron core. There is a 5 mm gap between the arrow tip and the magnet pole. The coil is energized by an alternating current of about 1 A (rms). The frequency of the current is controlled to within 0.01 Hz by a digital waveform generator. Since the magnetic force is attractive, independent of the sign of the current, the *excitation* frequency is *twice* the frequency of the current. The amplitude of the oscillating arrow is measured by slowly lowering a sensor (a 25 µm thick stainless steel reed, ('3'), until it touches the arrow when at maximum deflection. The electrically isolated arrow is connected to a 5 V power supply ('3') and contact between the reed and the arrow is detected electrically. The sensor is mounted on a micrometer stage; the amplitude is deduced from the micrometer reading when contact occurs. The measurements are reproducible to within a few hundredths of a millimeter. The data acquired with the driven sample arrow cover the resonance of the fundamental mode and are shown in Fig. 7.

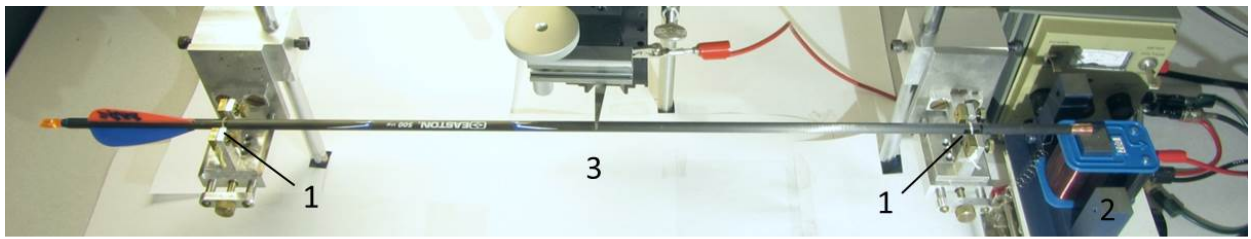

**Fig. 5** Experiment to measure the transverse oscillation of an arrow. The numbered items are identified in the text

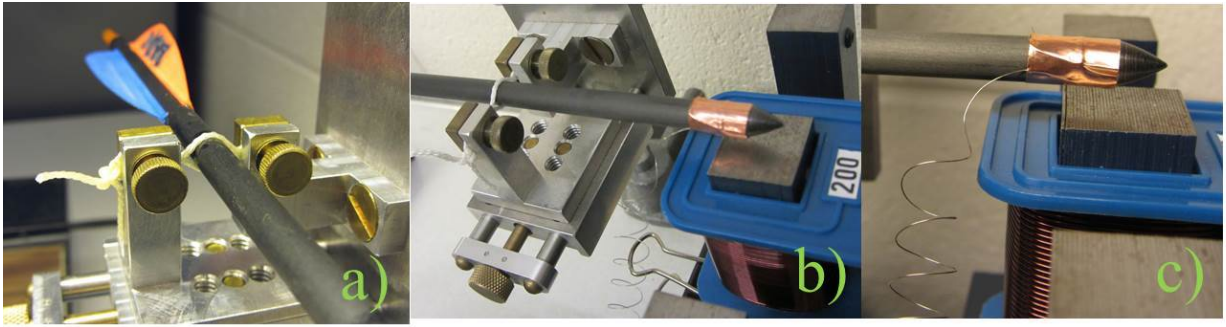

**Fig. 6** a), b) support at the nodes between two stretched strings: the vertical position is fixed but a tilt is allowed; c) the steel tip above the magnet. The electrical connection to the arrow is provided by the wire on the left

From the stiffness data in Fig. 3, we know that the restoring force is proportional to the displacement. This is the signature of a harmonic oscillator. The theoretical treatment of a driven harmonic oscillator yields the following equation for the mid-point amplitude $y_m$ as a function of the driving frequency $f$,

$$y_m(f) = \frac{y_m^0}{\sqrt{1+\left(2Q(1-f/f_0)\right)^2}} \quad , \tag{5}$$

where $f_0$ is the resonance frequency and $y_m^0$ is the amplitude on resonance. The '$Q$ factor' is proportional to the rate of energy loss due to friction. The curve in Fig. 7 has been calculated with Eq. (5), adjusting $f_0$ and $Q$ for best fit. The resulting resonance frequency is

$$f_{0,\exp} = (84.8 \pm 0.1)\,\text{Hz} \quad . \tag{6}$$

For the $Q$-factor we find $Q = 400 \pm 30$. If this parameter is large, the frictional energy loss in the system is low and the resonance is narrow. The damping time, $\tau_{1/e}$, during which a free oscillation decreases to $1/e = 0.37$ of its initial amplitude, is given by

$$\tau_{1/e} = \frac{Q}{\pi f_0} \quad . \tag{7}$$

With our values for $Q$ and $f_0$, the damping time amounts to $\tau_{1/e} \sim 1.5$ s. This is longer than the flight time of an arrow in most target archery situations. Thus, when the arrow hits the target, most of the initial transverse oscillation will still be present, if aerodynamic damping can be neglected.

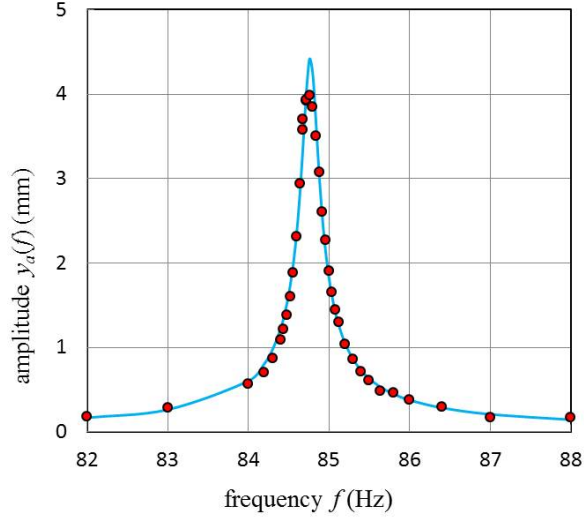

**Fig. 7** Transverse oscillation amplitude of the sample arrow versus excitation frequency (solid dots). The curve represents Eq. (5) with the parameters $f_0$ = 84.4 Hz and $Q$ = 400

### 2.3.2 Finite-element modelling

Advances in computing power and software have made it possible to simulate the dynamics of physical objects numerically. Of extensive use in this context is the finite-element (FE) method. This is a technique for finding approximate solutions to boundary-value problems represented by partial differential equations. To this aim, the structure under study is sub-divided into small volume elements. In each element the physics problem is solved exactly and then the elements are matched. The FE model used here is the 3D Structural Mechanics Module of the commercial software COMSOL [6] Similar computer modelling studies of arrows have been reported in the literature [7].

The arrow and its parts are defined in the program with some simplifications in shape but with masses that match those of their real counterparts exactly. Young's modulus $Y_z$ of the shaft material is chosen such that for the given moment of inertia $J_{xx}$, the measured stiffness is reproduced. Thus, all known static properties are represented in the FE model. Constraints and external forces are turned off, so the calculation models *free* motion.

Among the quantities that COMSOL calculates are the coordinates of the two nodes, $z_\alpha^C = 0.125$ m and $z_\beta^C = 0.630$ m and the frequency of the fundamental mode, $f_{0,C}$ = 85.4 Hz.

The excellent agreement between the calculated and the measured frequency, Eq.(6), suggests that one may use the model with confidence to calculate other parameters and to make predictions. For example, the model knows the kinetic energy for each element and thus the time-averaged kinetic oscillation energy of the entire object. The total energy $E_A^{osc}$ is the sum of potential and kinetic energy and, averaged over time, the two are equal. Using the fact that $E_A^{osc}$ is proportional to the square of the amplitude of the central displacement, $y_m$, we obtain $E_A^{osc} =$ 1080 J/m² $\cdot y_m^2$ for our arrow. For instance, with a relatively large amplitude of $y_m$ = 2 cm, the oscillation energy is $E_A^{osc} \sim 0.43$ J. We note that the oscillation energy is small compared to the kinetic energy of the moving arrow. This will be discussed further in sect. 3.3.2.

The FE model is also useful in predicting the effect of changes when considering modifications of a given arrow. For instance, when the length $L$ of the sample arrow is decreased by 5 cm (by 7%), the frequency $f_0$ increases to 96.0 Hz (by 13%). When the stiffness $S$ is increased by 20% (by lowering the 'spine' from 500 to 400), the frequency increases to 95.0 Hz (by 12%).

The FE code also calculates the higher resonant modes, which are probably unimportant in archery. For example, the second mode occurs at 272 Hz and has three nodes at $z$ = 0.070 m, 0.365 m, and 0.670 m.

### 2.3.3 Theory of vibrating beams

The Euler-Bernoulli theory [5], mentioned earlier, can be formulated to describe the *time-dependence* of the motion of beams. This leads to closed-form expressions for performance parameters, such as the resonance frequency, in terms of the properties of the system. Simplifying our arrow as a thin, uniform, free, vibrating beam of stiffness $S$, length $L$ and linear mass density $\mu'$, we arrive at the following equation of motion

$$\frac{\partial^4 y(z,t)}{\partial z^4} = -\frac{\mu'}{S}\frac{\partial^2 y(z,t)}{\partial t^2} \quad . \tag{8}$$

This equation relates the changes of the displacement from the axis, $y$, with position $z$ along the arrow and with time $t$. The length $L$ of the beam enters through boundary conditions. The solutions of Eq. (8) are discrete, associated with the modes of oscillation. They describe the shape of the bent beam (in particular the location of the nodes) and the frequency of a given mode. The result for the frequency of the fundamental mode is

$$f_{0,th} = \frac{1}{2\pi}\sqrt{\frac{S}{\mu'}}\left(\frac{c_0}{L}\right)^2 \quad . \tag{9}$$

The constant $c_0$ = 4.730 is the smallest (discrete) solution of the boundary equation $\cos(c) \cdot \cosh(c) = 1$. For the linear mass density $\mu'$ of the idealized beam, we average the arrow mass over its entire length, $\mu' = M / L$ ; values for $M$, $L$, and $S$ are from table 1. The result, $f_{0,th}$ = 97.2 Hz, differs from the measured value, $f_{0,exp}$ = (84.8 ± 0.1) Hz (Eq. (6)), by quite a bit. Nevertheless, Eq. (9) is still very informative because it tells us *how* the free oscillation frequency depends on length, stiffness and density. For example, if we wanted to increase the frequency, say, by 10%, we would have to decrease the arrow length by only 5%, but we would have to increase the stiffness by 20%. Moreover, we see that if a stiffness increase were accompanied by an increase of the linear density (which is likely to be the case), the two changes would tend to compensate each other.

## 3 THE BOW

The task of the bow is to store the mechanical energy that is produced by the archer. A good bow should, when triggered, transfer as much of the potential bow energy as possible into kinetic energy of the arrow.

Some components of the bow [8] are identified in Fig. 8. The rigid center part, made from cast aluminum, is the *riser*. Attached to it are the elastic *limbs*, which bend when the bow is drawn and store energy. The *grip* is where the hand of the archer pushes while shooting. The point on the *string* where the arrow nock is placed in preparation for the shot is the *nocking point*. Typical for a compound bow are the extra *cables* and the *cams* (off-center pulleys).

### 3.1 Measuring the stored energy

The setup to measure the draw force is shown in Fig. 8. The bow is tied down at the grip and pulled up by the force $F_d$, applied at the nocking point in the direction indicated at the top of the figure. This 'draw force' is a function of the distance $s$ between the grip and the nocking point. For a relaxed bow this distance equals $s_0$, called *brace height*. While the bow is drawn, the nocking point moves by a distance $s_D$, and at full draw $s = s_0 + s_D$, the *true draw length*. For our sample bow [8] $s_0 = 0.185$ m and $s_D = 0.481$ m.

During the measurement, the bow is supported just at the grip and at the nocking point and can rotate around a straight line through these two points. The wooden bar on the left is there to prevent this. The force is applied by a rope attached to the nocking point. A block and tackle (not shown) is used to help pulling the rope. The force is measured by a piezo-electric *load cell* and is displayed on a read-out at the bottom of the picture.

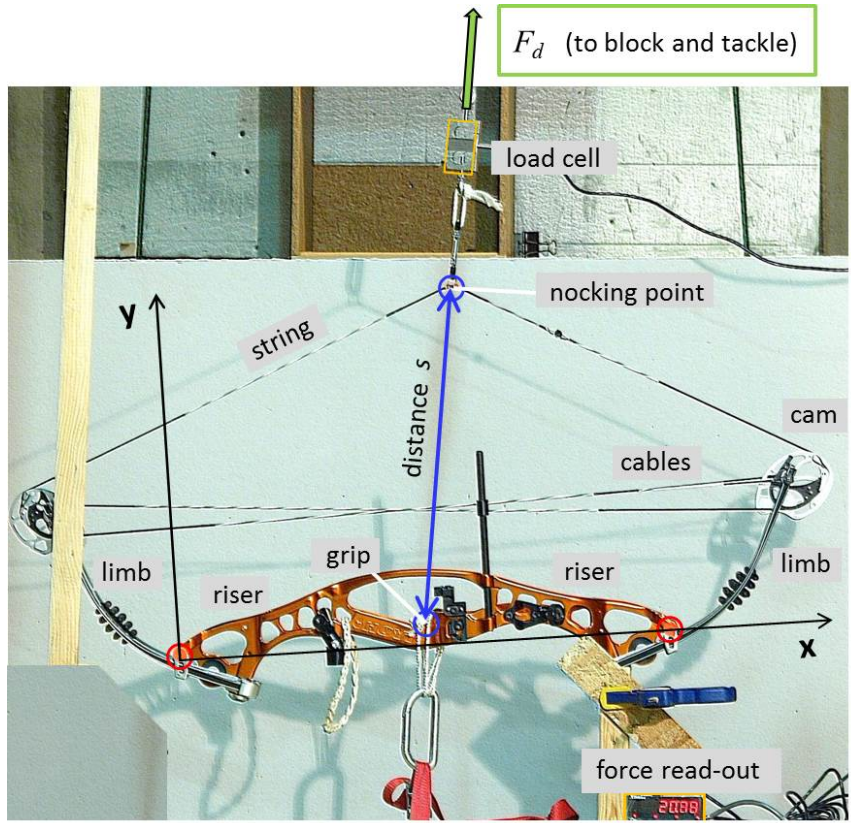

**Fig. 8** Definition of some bow terms and setup to measure the draw force as a function of the distance *s*. In this picture, *s* = 0.432 m

To determine the amount of stored energy, we need to know the draw force $F_d$ as a function of the drawn distance *s*. To measure this we draw the bow in small steps, taking a picture after every step. To reduce parallax, a 200 mm lens at 10 m is used. After some 15 exposures, having reached full draw, this process is reversed, until back to the rest position. Consequently, data are obtained when *drawing* the bow ($F_d^{in}(s)$) and when *releasing* it ($F_d^{out}(s)$).

The draw length *s* is extracted by digitizing [9] the coordinates of the points of interest in each picture. A bow-fixed coordinate frame with an absolute length scale is defined, based on the two reference points on the riser, marked by red circles.

The data are shown in Fig. 9. It is typical for a compound bow that the draw force first rises steeply, then is roughly constant over much of the draw and then decreases sharply. In the present case this 'let-off' at full draw is about 65 %. The sharp increase beyond full draw occurs because the cables run out of length. The same graph would look much different for a *conventional* bow, where the draw force increases monotonically over the entire draw.

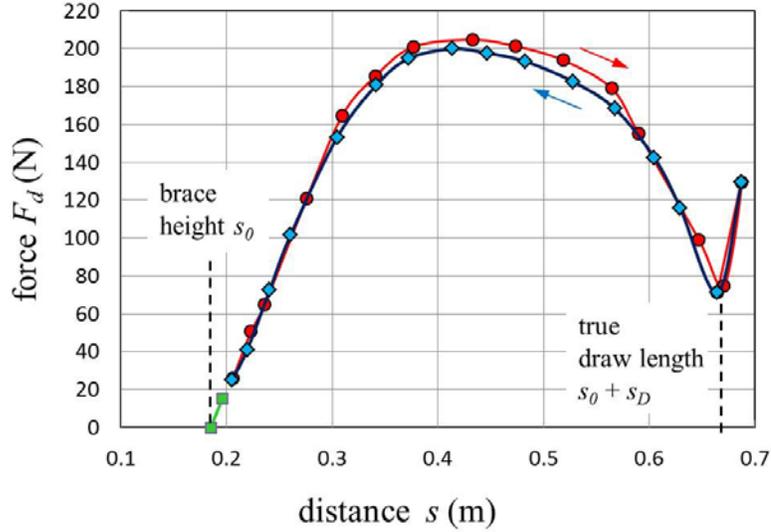

**Fig. 9** Draw force versus draw distance. The circles mark data taken while pulling the string ($F_d^{in}$) and the diamonds ($F_d^{out}$) while releasing it. The bow grip is at $s = 0$. The brace height $s_0 = 0.185$ m and the true draw length $s_0 + s_D = 0.665$ m are marked by dashed lines. The curves are spline fits to the data. The two squares near $s_0$ are described in sect. 3.2

The data collected during the force measurement are also valuable in other ways. For instance, they can be used to verify that the nocking point during release travels on a straight line. This is necessary to keep the force on the arrow pointed in a fixed direction and is a consequence of properly matched limbs.

The energy $W_B^{in}$ transferred to the bow by the archer is the work done on the bow and may be calculated by integrating $F_d^{in}(s)$ from brace height to full draw,

$$W_B^{in} = \int_{s_0}^{s_0+s_D} F_d^{in}(s)\,ds \quad . \qquad (10)$$

By numerically integrating a spline fit to the data one obtains for this *input* energy to the bow $W_B^{in} = (73.7 \pm 0.5)\,\text{J}$. The quoted uncertainty is based on a ±3 N uncertainty of the force measurement.

The energy *available for the shot* is obtained in an analogous way by integrating over $F_d^{out}(s)$ (diamonds in Fig. 9). The resulting *released* energy equals $W_B^{out} = (71.0 \pm 0.5)\,\text{J}$. This is smaller than $W_B^{in}$ by 3 to 4%. This difference arises because some energy is dissipated by internal friction in the limbs and by mechanical friction in the cams.

In the days of wooden bows, frictional loss of up to 20% was reported [2]. Because of the importance of internal friction in strained wood, presumably because fibers are rubbing against each other, the choice of wood was crucial in classic bow making.

## 3.2 String tension and string elasticity

The string tension is of interest because the length of the string depends on it and the string may thus contribute to the dynamics of the bow. One of the two squares in Fig. 9 marks the brace height. The other is measured by mounting the bow with the string horizontal and observing the string displacement with a known small weight dangling from the nocking point. This determines the *slope* of the force curve at brace height, $(dF_d/ds)_{s_0} = (1320 \pm 20)$ N/m. This quantity is related to the string tension $T_0$ of the relaxed bow by $T_0 = ¼\, \ell\, (dF_d/ds)_{s_0}$, where $\ell = 0.930$ m is the free string length between the cams. The result is $T_0 = (307 \pm 5)$ N.

As the bow is drawn, the string tension *decreases* until, at full draw, it is only about 50 N. This information is extracted from pictures like Fig. 8 (taken for the draw force measurement in sect. 3.1) using the fact that the three force vectors acting at the nocking point (two string tensions and the known draw force) must add to zero.

The elasticity of a string is governed by Hooke's law from which follows that a change $\Delta T$ in string tension causes the string to elongate by $\Delta\ell = \chi\cdot\ell\cdot\Delta T$. In order to determine the constant $\chi$ for the string of the sample bow we measure its elongation (to $\pm 0.5$ mm) when dangling a weight from it. To produce a significant effect, one needs a large force. Here, we use a weight of 2000 N and a crane. By lowering this weight gradually onto an industrial scale, the applied string tension (the difference between the weight and the scale reading) is varied from 100 N to 2000 N to test the linearity of $\Delta\ell(T)$. The result of this experiment is $\chi_{exp} = (5.6 \pm 0.3)\cdot 10^{-6}$ N$^{-1}$.

From this, we find that while the string tension changes by $\Delta T = -250$ N during the draw, the string *contracts* by a calculated 0.1−0.2 %. It seems that string elasticity is not important.

However, it is interesting to compare our $\chi_{exp}$ measurement with what one might expect. Even though the composition of our string is proprietary, it is likely to consist to a large part of the ultra-high-molecular-weight polyethylene Dyneema, for which the density equals $\rho_{Dy} = 975$ kg/m$^3$ and Young's modulus equals $Y_{Dy} = 110$ GPa. The constant $\chi$ is then given by $\chi_{Dy} = \rho_{Dy} / (\mu_{string}\, Y_{Dy})$, where the linear mass density of the string in use (without the serving) is $\mu_{string} = 3.3$ g/m. Inserting numbers yields $\chi_{Dy} = 2.7\cdot 10^{-6}$ N$^{-1}$, in reasonable agreement with our measurement.

## 3.3 Partition of the available bow energy

### 3.3.1 Arrow and bow kinetic energies

The kinetic energy of the arrow after it has left the string equals

$$E_A = \frac{1}{2} M v_0^2 = (57.1 \pm 0.8) \text{ J} \quad . \tag{11}$$

The arrow mass $M$ is listed in table 1, and the measurement of the launch velocity $v_0$ will be discussed in Sect. 5.1.

The kinetic energy of the arrow is quite a bit smaller than the available potential energy $W_B^{out} = (71.0 \pm 0.5)$ J. This is mostly because, at the instant when the arrow separates from the string, various parts of the bow are also moving more or less rapidly and represent a bow kinetic energy $E_B$. Klopsteg proposed [2] the useful concept of 'virtual' bow mass $\tilde{m}$, defined by the equation $E_B \equiv \tilde{m} v_0^2 / 2$, where $v_0$ is the launch velocity. In effect, $\tilde{m}$ corresponds to the average of the masses of all parts of the bow, weighted with the square of their individual velocities.

Let us assume that the *entire* available potential energy is divided up between the kinetic energies of the bow and the arrow, or $W_B^{out} = E_A + E_B$ (we will revisit this assumption in the next section). It is easy to see that then the virtual bow mass equals

$$\tilde{m} = \frac{2 W_B^{out}}{v_o^2} - M \quad . \tag{12}$$

Inserting the measured values for $W_B^{out}$, $v_0$ and $M$, we obtain $\tilde{m} \sim 5$ g.

The main contributors to $\tilde{m}$ are the fast moving parts of the bow. The string, for instance, , contributes 1.1 g to $\tilde{m}$ (setting its linear density to $\mu_{string}$ = 3.3 g/m), and the serving on the string another 0.7 g. The D-loop (a piece of string, installed at the nocking point to connect a release aid) and the peep sight (lodged between strands of the bow string and a part of the aiming arrangement) together add 0.7 g. So far, this accounts for half of the virtual bow mass and it is plausible that the other half is due to other parts of the bow, such as the limbs and the cams, which are more massive but are moving more slowly.

Re-arranging Eq.(12) leads to an expression for the arrow velocity at launch:

$$v_0 = \sqrt{\frac{2 W_B^{out}}{M + \tilde{m}}} \quad . \tag{13}$$

This equation explains why lighter arrows are faster and also why using a lighter bow string (smaller $\tilde{m}$) increases the arrow speed. Figure 10 shows the measured launch speed of three arrows with different masses, loosed with the sample bow, together with $v_0$ calculated using Eq. (13).

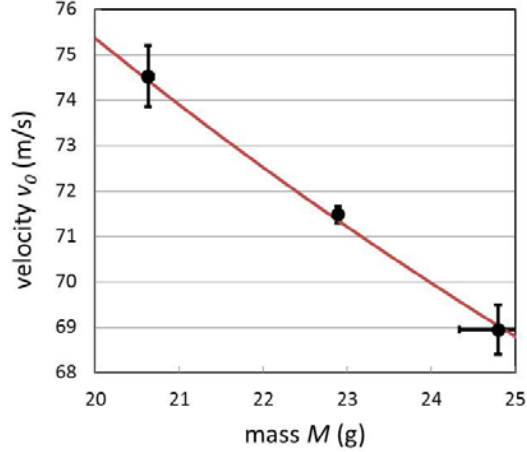

**Fig. 10** Launch velocity for three arrows of different mass using the same bow, measured with a chronograph [14]. The solid line is calculated from Eq. (13) with $W_B^{out}$ = 71.0 J and $\tilde{m}$ = 5 g

Let us define the *efficiency* $\varepsilon$ of a bow as the fraction of the *available* stored energy that is transferred to the arrow. Then

$$\varepsilon \equiv \frac{E_A}{W_B^{out}} = \frac{M}{M + \tilde{m}} \quad . \tag{14}$$

This remarkably simple expression states that the same bow, used with a heavier arrow, is *more* efficient, and that it must be one of the goals of bow design to minimize the virtual mass $\tilde{m}$. With $M$ = 20.6 g and $\tilde{m}$ = 5 g the efficiency of our bow turns out to be $\varepsilon$ = 0.80.

### 3.3.2 Other forms of energy

In the previous section we have postulated that, at launch, *all* potential energy is converted to the combined kinetic energy $E_A + E_B$ of the arrow and the bow. This is not exactly true, since other forms of energy are present as well. In the following, we list some of these contributions and argue that they are unimportant, compared to the available energy $W_B^{out}$.

'De-nocking' energy. During acceleration, the arrow is clipped onto the bow string by a plastic nock. It takes a certain force, $F_N$, to pull the arrow off the string. One can measure this force by dangling a nocked arrow vertically from the string, gradually adding weight to it until it falls off. Typically, one finds $F_N \sim$ 10 N.

At the end of the acceleration, the string passes through brace height and continues with the arrow still nocked to it. Now there is a *decelerating* force on the arrow, which grows until that force has reached $-F_N$ and the nock is pulled off the string. From the derivative of the draw force near brace height, sect. 3.2, we conclude that $F_N$ will be reached about 1 cm beyond brace height. The work done by the average decelerating force (½$F_N$), acting over that distance, equals 0.05 J, which is negligible.

Recoil energy. After the launch, bow and arrow move in opposite directions and conservation of momentum requires that $M \cdot v_0 = M_B \cdot v_B$, where $M_B = 6.5$ kg is the mass of the bow and $v_B$ its speed just after separation from the arrow. The bow recoil energy is then $E_{recoil} = E_A(M/M_B)$. Since $M/M_B$ is about 0.003, the recoil energy is ~ 0.18 J and may be ignored.

Today's technology offers the possibility to observe the motion of the bow directly. Attaching a wireless accelerometer [10] to the bow, we observe an initial acceleration spike of about 37 m/s$^2$ during 6 to 8 ms, or a recoil velocity of $v_B$ = 0.26±0.04 m/s. This amounts to $E_{recoil} = M_B v_B^2/2$ =(0.23 ± 0.06) J. This convincingly verifies the theoretical reasoning in the previous paragraph.

Arrow oscillation energy. Transverse arrow oscillation is discussed in sect. 2.3.2, where it is found that even with a relatively large central displacement amplitude of $y_m$ = 2 cm, the oscillation energy equals $E_A^{osc}$ ~0.43 J, and may be ignored.

Potential energy stored in the bow string. The potential energy stored in a string is proportional to Hooke's constant. In sect. 3.2, we discuss the change of string tension $\Delta T = -250$ N during the draw and find that a 1 m long bow string *contracts* by 1 – 2 mm. The corresponding loss of potential energy is 0.2 to 0.3 J, which may be ignored.

## 4  INTERNAL BALLISTICS

### 4.1  Acceleration

The arrow is accelerated over a distance $s_D$, from full-draw to brace height, by the force $F_d^{out}$, discussed in sect. 3.1. This force varies with draw distance and has a maximum $F_{d,max}^{out}$ of about 200 N. The corresponding acceleration, $F_{d,max}^{out}/M$, is about 1000 times earth gravity. It is interesting that on-board accelerometers that are built into the arrow tip [11] can be made to tolerate such an acceleration.

The acceleration phase lasts only about $\Delta t \approx 2s_D/v_0 \approx 13$ ms. This estimate agrees well with the observed 12 – 14 ms for the duration of the noise of the accelerating string between point '1' and '2' of the acoustical record shown in Fig. 12.

### 4.2  Arrow buckling

During acceleration, the arrow is subjected to an axial force $P$, compressing it lengthwise between nock and tip. To estimate this axial load, we note that the entire arrow is accelerated by the force $F_d^{out}$, shown in Fig. 9. The fraction that is needed to accelerate *just the tip* is then $(m_{tip}/M)F_d^{out} \approx 0.3 F_d^{out}$. This force, which is acting at the nock, is transmitted through the shaft to the tip and thus represents the axial load. The largest load encountered during acceleration is then $P_{max} \approx 0.3 F_{d,max}^{out} \approx 60$ N.

Some 250 years ago, L. Euler derived the following formula for the maximum load $P_{crit}$ that a long slender column can support before it exhibits the so-called buckling instability and collapses.

$$P_{crit} = \frac{\pi^2 S}{L^2} \quad . \tag{15}$$

Inserting the values for the stiffness and arrow length for the sample arrow from table 1, we find $P_{crit} \approx 100$ N. The estimated axial load $P_{max}$ for our arrow is thus a bit more than half the critical load for Euler buckling, probably representing an acceptable safety margin. The following changes would bring the axial load closer to the critical value: increasing either the bow draw weight, the tip mass, or the arrow length, or decreasing the stiffness (i.e., increasing *Spine*). For instance, based on the above, one would expect arrows with *Spine* in excess of 800 to fail.

### 4.3 Archer's paradox

This term was coined in 1913 when people started to wonder how the arrow gets around a bow which is in fact blocking a straight path from the nock at full draw to the target [12]. This puzzle has since been explained by the fact that, given the right conditions, the arrow undulates horizontally *around* the bow.

Modern bows commonly feature center-shot risers with an unobstructed path for the arrow. In addition, when a release aid is used, the transverse oscillation tends to be smaller than when the string is released from the finger tips. So, for our sample bow, the Archer's Paradox is really not much of an issue. However, it seems that a paper on archery without a treatment of the Archer's Paradox is incomplete and since we have acquired all the necessary tools to *calculate* the transverse motion of an arrow during acceleration, we give it a try.

Figure 11a shows side views of a bow at rest and when fully drawn. The top view on an exaggerated transverse scale is shown in Fig. 11b. The bow which obstructs the line to the target (horizontal dashed line) is depicted by a rectangle at $z = 0.6$ m. The arrow labelled '1' has just been launched ($t = 0$). When it is released by the fingers of the (right) hand, its tail end (circle) is propelled in the +x-direction, along the dashed line, and a transverse oscillation is excited.

The Euler-Bernoulli beam theory (sect. 2.3.3) allows us to calculate the transverse displacement anywhere along the arrow, at any time $t$. We also know the location of the arrow at time $t$ because we know its acceleration. Thus, we can calculate the snapshots '2' (as the mid-point of the arrow nears the bow) and '3' (after the arrow has separated from the string and the tail end has cleared the bow). The center of mass of the arrow, of course, travels on a straight line parallel to the z-axis.

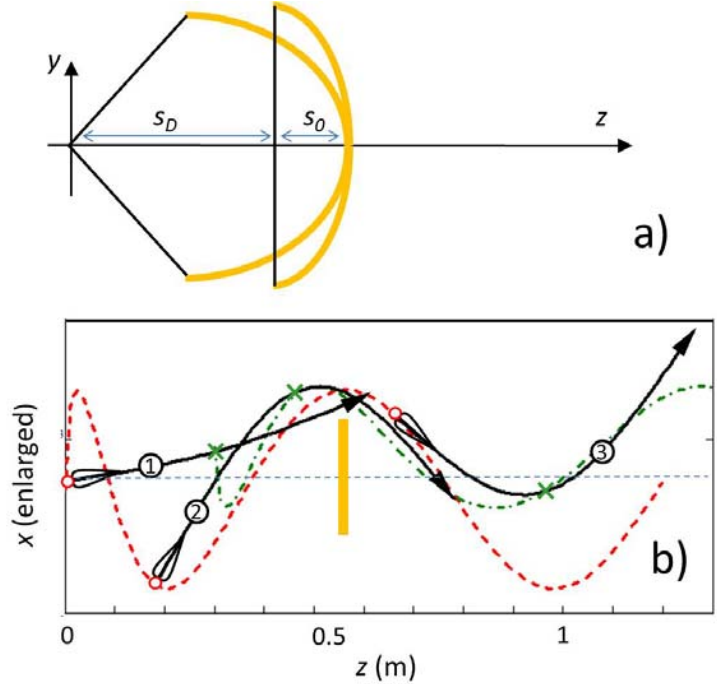

**Fig. 11** a) Side view of the bow, drawn and relaxed. b) Top view (the transverse displacement is exaggerated): three arrow positions during acceleration. Also shown is the trace of the nock (circle, dashed line) and of the midpoint of the arrow (cross, dash-dot line). The rectangle at $z = 0.6$ m represents the bow which is blocking the shooting direction

The trace of the tail end (circle) is shown as a dashed line and that of the midpoint (cross) as a dash-dot line. It is apparent that both these points pass the bow with good clearance, in other words, the arrow 'wiggles around' the bow. Such behavior requires the interplay of velocity, arrow length and transverse frequency. For the present calculation we use $v_0 = 71$ m/s, $L = 0.6$ m and $f = 85$ Hz. It is interesting to compare Fig. 11 with a slow-motion movie of an arrow being launched [13].

## 5 EXTERNAL BALLISTICS

### 5.1 Launch velocity

The usual method to measure the speed of an arrow is by the use of a chronograph in which the projectile successively interrupts two light barriers about 30 cm apart. The speed is deduced with an accuracy of about 0.5 m/s from the measured time difference. Several affordable chronographs are on the market; here we use the 'Chrony' [14]. Averaging ten measurements at $Z = 2$ m (practically in front of the bow) yields the launch velocity for our sample arrow

$$v_0 = (74.5 \pm 0.2) \text{ m/s} \quad . \tag{16}$$

## 5.2 Deceleration

Aerodynamic drag causes a decelerating force opposite to the velocity $v$. At speeds $v$ larger than about 10 m/s, this force is known to be proportional to the *square* of the velocity

$$\frac{dv}{dt} = -\kappa v^2 \quad , \tag{17}$$

where $\kappa$ [m$^{-1}$] is the *velocity decay rate*. Solving this equation for motion in the $z$-direction, shows that the arrow velocity decreases *linearly* with distance provided that $\kappa \cdot z \ll 1$.

$$v(z) = v_0 \exp(-\kappa z) \approx v_0 (1 - \kappa z) \quad . \tag{18}$$

In order to measure the velocity decay rate for our sample arrow we need velocity data for long shooting distances. Because this would put a chronograph with its ~20 cm opening at risk, we have devised an alternative method to measure the arrow velocity, which makes use of the shooting noise. It requires microphones, a computer and audio recording and analysis software with milli-second resolution. The latter is available as freeware (e.g., [15]).

A typical sound recording of a shot is shown in Fig. 12. One can clearly identify the noise of the accelerating string between '1' and '2', followed by the ring-down of the oscillating string after the arrow has left. Later, the sound from hitting the target is seen at '3'. The time between '2' and '3' equals

$$\tau = \frac{Z}{\bar{v}} - \frac{\Delta}{c} \quad , \tag{19}$$

where $Z$ is the target distance, $\bar{v}$ the average speed of the arrow, $c$ the speed of sound and $\Delta$ the distance from the microphone to the bow, minus the distance from the microphone to the target.

We are using a shooting machine to eliminate the human factor and to shoot arrows at targets between $Z = 27.4$ m and 82.3 m. The use of *two* microphones results in two independent measurements and allows one to eliminate the speed of sound $c$ from the analysis. These microphones are installed at 4.6 m and 36.6 m from the bow and remain fixed for the entire experiment.

A total of 25 shots have been loosed at 7 distances. For each shot, $\Delta_1$ and $\Delta_2$ for the two microphones is determined, the noise 'start' times are read from the sound tracks, and $\bar{v}$ is deduced. The uncertainties are based on a distance error of 30 cm and a timing error of 1 ms. The results are shown as black dots in fig. 13.

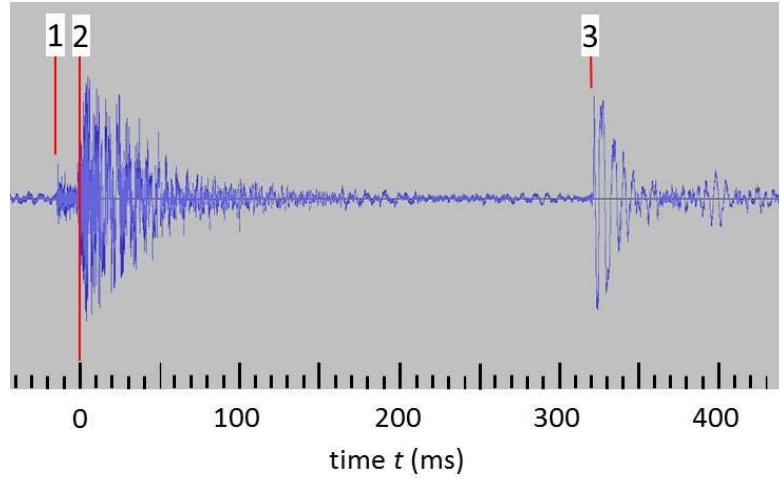

**Fig. 12** Sound recording of a shot at 18.3 m using the program Audacity [15]. The string is released at '1', the arrow leaves the string at '2', and '3' marks the noise from hitting the target.

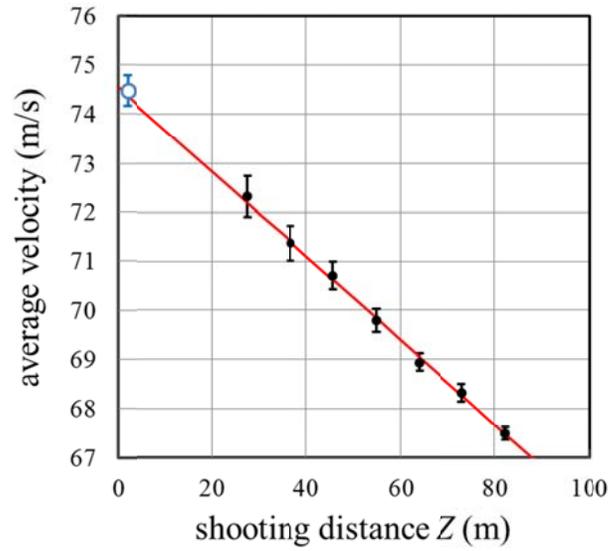

**Fig. 13** Velocity $\bar{v}(Z)$, averaged over the shooting distance, extracted from sound recordings (dots) and the instantaneous velocity at 2 m measured with a chronometer (open circle). The solid line corresponds to Eq.(20) with $\kappa = 0.00231$ m$^{-1}$

From Eq.(18) we expect for the *average* velocity $\bar{v}(Z)$ between bow and target

$$\bar{v}(Z) \approx v_0 \left(1 - \tfrac{1}{2}\kappa Z\right) \quad . \tag{20}$$

Fitting this equation to the data in Fig. 13 (solid line) results in the following velocity decay rate

$$\kappa = (2.31 \pm 0.16)\,10^{-3}\,\text{m}^{-1}. \tag{21}$$

## 5.3 The drag coefficient

One expects the drag force to be proportional to the cross sectional area $\pi R^2$ of the arrow shaft and to the air density $\rho_{air}$, which itself depends on temperature, barometric pressure and altitude. Separating these known factors from the unknown remainder leads to the definition of the dimensionless *drag coefficient* $C_D$:

$$\kappa \equiv \frac{\pi R^2 \rho_{air}}{2M} C_D \quad . \tag{22}$$

Inserting $R$ and $M$ from Table 1, $\rho_{air} = 1.19 \pm 0.01$ kg/m$^3$ (at 23ºC) and the measured $\kappa$, Eq. (21), we obtain for the drag coefficient of our sample arrow $C_D = 1.94 \pm 0.14$.

The drag coefficient $C_D$ is affected by details of the arrow, like its fins and its surface roughness, and whether the aerodynamic flow is laminar or turbulent. The latter is assessed by Reynold's number $Re = 2R\rho_{air}v_0/\nu_{air}$, where the viscosity of air at 23°C is $\nu_{air} = 18.5 \cdot 10^{-6}$ Pa·s. Our sample arrow at launch velocity has a Reynolds number of about 35'000, indicating turbulent flow.

The aerodynamics of arrows was first studied in the 1930s [16], including wind tunnel measurements (J. Rheingans [3] p.236). It was found that the fletching contributes about 40% to the drag force. More recent studies of $C_D$ for a variety of arrows around $Re \approx 25'000$ suggest [17] that $C_D$ is typically between 1.5 and 2.5. Measurements in a wind tunnel [17,18] are consistently lower than those with arrows shot by archers, because the wind tunnel conditions do not replicate those of a real, rotating and oscillating arrow in flight. It is worth noting that Ref. 18 confirms that the fletching contributes about 35% to $C_D$.

## 5.4 Trajectory and sight adjustment

Let us study the flight of the arrow in a frame where the $z$-axis is a horizontal straight line from the nocking point to the center of the target. The arrow is launched at velocity $v_0$, at an angle $\theta_0$ with respect to this $z$-axis. It is affected by a drag force opposite to its velocity and a gravitational force pointing straight down ($-y$-direction). It is conceivable that there is also *lift*, an upward force, perpendicular to the drag force. There is no evidence for such a force, as we shall see.

Mathematically, the resulting trajectory is described by a pair of coupled differential equations. These equations can be analyzed numerically, for instance using the so-called Runge-Kutta method, as described in Ref. [19]. A solution in *closed* form (where the result is a formula) is attainable when one neglects drag in the vertical direction [20]. This is a good approximation when shooting at targets closer than about 100 m, because then the launch angle $\theta_0$ is small and the vertical velocity is always less than $v_0 \sin\theta_0$.

Ballistic equations are treated extensively in the literature [19,20] Here, we concentrate on adjustments to the bow sight and are therefore interested in the launch angle $\theta_0$ required to hit a target on the z-axis at a distance $z = Z$. Solving the approximate equations [20] leads to

$$\theta_0(Z) = \frac{1}{2}\arcsin\left(\frac{gZ}{v_0^2}\left(1+\sum_{n=1}^{\infty}\frac{(\kappa Z)^n}{(n+1)!}\right)\right) \quad , \tag{23}$$

were $\theta_0$ is in radians and $g$ equals the earth gravitational acceleration. The effect of drag is represented by the sum, which arises from the series expansion of an exponential function. It is easy to see that only one or two terms are of practical importance. The value for the velocity decay rate $\kappa$ is taken from Eq. (21). Figure 14 shows three calculations of the launch angle as a function of distance: the exact Runge-Kutta calculation, the approximate result according to Eq. (23) and the same without drag ($\kappa = 0$ in Eq. (23)).

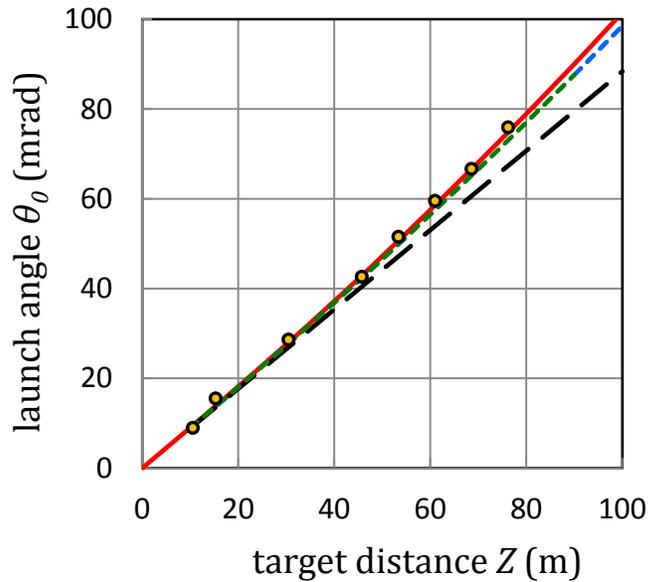

**Fig. 14** Launch angle versus target distance. The solid line depicts the numerical solution of the ballistic equations with drag, using a 4[th] order Runge-Kutta algorithm. The approximate treatment of drag is shown by short dashes. If there were no aerodynamic drag, the long-dashed line would result. The data points are the result of a measurement described in the text

Usually, aiming with a bow is accomplished by establishing a line of sight containing the eye of the shooter, the aiming point of the bow sight and the center of the target. Often, the eye is aided by a *peep sight*, a small plastic ring, wedged between the strands of the bow string, directly in front of the archer's eye.

The bow sight establishes a reference point in front of the bow, for instance by means of a pin on some support, fixed to the bow. The sight is adjustable vertically. In target archery, the position of the sight is read out on an attached length scale. It is assumed that the bow is fixed to the line of sight (this, in fact, is accomplished by the skill of the archer). So, moving the sight up and down changes the launch angle. It is a matter of geometry to convert the sight scale reading and the distance between the peep and the sight to an angle, which, within a constant offset, equals the launch angle $\theta_{0,exp}$. The offset is determined from the condition that the launch angle must tend to zero as the distance $Z$ gets smaller. One must take into account that the sight is usually some 5 to 10 cm *above* the *z*-axis. Thus, there is a small angle between the line of sight and the *z*-axis. This leads to a correction which is significant for shooting distances less than about 30 m.

We are now in a position to *measure* the launch angle $\theta_0(Z)$. To this effect, we shoot at targets at various distances. For each distance we vary the sight position until we hit the center of the gold. Since no help from an Olympic archer is available, we use a shooting machine. These empirically determined sight positions, converted to a launch angle $\theta_{0,exp}$ are shown as points in Fig. 14. The agreement between the measurement and the Runge-Kutta calculation is surprisingly good. It supports the initial assumption that *no aerodynamic lift* is needed to explain our empirical findings.

Ballistic theory does not replace careful shooting-in of the bow but is does provide insight. In particular, the explicit expression for the launch angle $\theta_0(Z)$, Eq.(23), makes it possible to study how to change the bow sight in response to a change in launch velocity, $\Delta v_0$, aerodynamic friction, $\Delta \kappa$, or the shooting distance, $\Delta Z$. To this effect, we calculate the vertical shift $\Delta y(Z)$ of the target spot caused by such changes.

To simplify this task, we modify Eq. (23) by replacing the small $sin(\theta_0)$ by $\theta_0$, and by retaining only the first term in the sum (for $Z = 70$ m the second term is 20 times smaller). Then we calculate the differential $\Delta \theta_0$, omitting all terms $(\kappa \cdot Z)$ which contribute only little. The vertical shift on target, $\Delta y(Z) = Z \cdot \Delta \theta_0$, is then

$$\Delta y(Z) \approx Z \cdot \theta_0(Z) \cdot \left( \frac{\Delta Z}{Z} - 2 \frac{\Delta v_0}{v_0} + \frac{Z \cdot \Delta \kappa}{2} \right) \quad . \tag{24}$$

For instance, when shooting at $Z = 70$ m ($\theta_0 = 0.068$ radians), we find that a change of the shooting distance by 1 m, everything else being the same, would result in a vertical shift on target by ~7cm. An increase in launch velocity by 1 m/s would raise the target spot by ~13cm and a 20% change in the velocity decay rate $\kappa$ or the drag coefficient shifts the target spot by ~ 7 cm.

## 6  CONCLUSIONS

We apply scientific methods to topics ranging from the mechanics of bows and arrows, to the launch process and the flight of the arrow. These ideas are put to the test by a number of measurements carried out with a specific compound bow and arrow combination. The quantitative information that results from these experiments confirms the underlying physics laws, helps us understand our equipment and allows us to make predictions beyond the range of actual observations. Most of the described experiments require only modest tools and readers may find it interesting to acquire analogous results with their own equipment.

Physics is an exact science, but archery, as a whole, is not. The shooting of an arrow has many aspects that are too complicated to be easily attributed to a dominant variable, or it may be difficult to devise measurements that link a hypothesis with an observable phenomenon. Consequently, there are many topics in archery that lend themselves to endless speculation, for instance the mysterious concept of dynamic spine, or the relation between FOC and lift, or the effect of non-straightness of an oscillating and rotating arrow on its flight. It might be worthwhile to devote future research to some of these topics.

## ACKNOWLEDGEMENTS

The author wishes to thank P.L. Childress, T.M. Sampson and G. Wood and P.T. Smith for their interest in this work and for help with the experimental equipment, A.R. Dzierba for challenging discussions and constructive criticism, and J.C. Long and R.E. Noel for carefully reading the manuscript and for making valuable suggestions.